\documentclass[a4paper]{jpconf}
\usepackage{graphicx,color,amsmath,amssymb}
	
\begin{document}
		\title{Prospects for transient gravitational waves at $r$-mode frequencies associated
			 with pulsar glitches}
		\author{I Santiago-Prieto$^1$, I S Heng $^1$, D I Jones$^{2}$ and J Clark$^{3,}$$^ 4$}
		\address{$^1$ School of Physics and Astronomy, University of Glasgow, Glasgow
				 G12 8QQ, UK}
		\address{$^2$ Department of Mathematics, University of Southampton,
			 	 Southampton SO17 1BJ, UK}
		\address{$^ 3$ School of Physics and Astronomy, Cardiff University, 5 The Parade,
			 	Cardiff CF24 3AA, UK}
		\address{$^4$ Physics Department, University of Massachusetts, Amherst,
			 	Massachusetts 01003, USA}
		\ead{ignacio.santiago-prieto@ligo.org}
		%
		%%%%%%%%%% ABSTRACT  %%%%%%%%%%
		%
		\begin{abstract}
			Glitches in pulsars are likely to trigger oscillation modes in the fluid interior of 
			neutron stars. We examined these oscillations specifically at $r$-mode frequencies.
			The excited $r$-modes will emit gravitational waves and can have long damping 
			time scales $\mathcal{O}$(minutes - days). We use simple estimates of how much
			energy the glitch might put into the $r$-mode and assess the detectability of the 
			emitted gravitational waves with future interferometers.
		\end{abstract}
		%
		%%%%%%%%%% SECTION 1 %%%%%%%%%%
		%
		\section{Introduction}\label{sec:intro}
            	One possible source of burst-like gravitational waves is a {\it pulsar glitch}.
            	These are anomalies in the observed spin frequency of a neutron star (NS)
            	wherein there is a sudden increase in its rotational frequency followed by
            	an exponential recovery to nearly the pre-glitch rotation rate on timescales
            	of a few days--weeks. In most cases, these glitches are produced by young
            	pulsars.  It is believed  that these glitches are a consequence of global 
            	events within the star's fluid interior and its solid crust~\cite{lg06}. These 
            	glitches could excite a variety of quasi-normal modes \cite{dynam_glitches}.
            	A previous search for gravitational waves associated with a glitch in the Vela
            	pulsar in 2005 focussed on $f$-modes whose frequencies are believed to lie
            	in the range 1--3\,kHz~\cite{Velapaper2011} and last ${\mathcal O}$(100\,ms).
            
            	A possibility to be considered in gravitational wave searches of this type is the
           	excitation of $r$-modes oscillations during a pulsar glitch \cite{Rezania_crab}. 
            	These $r$-modes, which are a type of inertial modes (oscillation modes restored
            	by the Coriolis force), have frequencies proportional to the star's angular velocity
		and they are considered relatively more efficient GW emitters compared to other
		inertial modes \cite{LandF_where_are}. Also, these modes evolve in time with an
		exponential decrement that depends on the mode's frequency and dissipation
		processes \cite{Owenetal1998}. Reference \cite{Glam_And_rmodes_glitch}  
		details that when a rotational lag between the fluid component and the crust reach
		a critical value, an instability that excite r-mode oscillations within the NS interior 
		sets in. Because of this lag, the fluid component (already `pinned' to the crust by a
		series of vortices) will be unpinned and will transfer angular momentum to the crust
		which will consequently spin up.  As a result of this process a pulsar glitch will be
		produced. In that work, the authors considered short-wavelengths modes at which
		the quadrupole moment $l=m=2$ would not be excited.
		
            	In section~\ref{sec:signal} we describe the form and duration of the GW signal 
		one might expect from $r$-mode oscillations. In section~\ref{sec:plausibility}, 
		we explore the energetics of the $r$-mode signals associated with pulsar glitches
		and the plausibility of detecting GWs from $r$-modes. Finally, in section~\ref{sec:Discussion},
		we summarise our findings.
		%
		%%%%%%%%%% SECTION 2 %%%%%%%%%%
		%
		\section{Gravitational waves signals from $r$-modes in neutron stars}\label{sec:signal}

            	We parameterise the expected GW signal from an $r$-mode oscillation
            	in terms of harmonic ringdown gravitational waves signals \cite{burstjointsearch},
			%
			%%%%%%%%%% signal equation %%%%%%%%%%
			%
			\begin{eqnarray}
				 h_{+} (t) = h_0 \cos(\omega_r t) e^{-t/\tau_r}				 
				 && 
				 h_{\times} (t) = h_0 \sin(\omega_r t) e^{-t/\tau_r},
				\label{eq:signal}
	 		\end{eqnarray}
           	where $h_{+}$ and $h_{\times}$ are the gravitational wave signal polarisation `plus' (+)
            	and `cross' ($\times$), $\omega_r=2\pi f_r$ is the frequency of the $r$-mode oscillation, 
            	$\tau_r$ is the damping time of the signal and $h_0$ is the amplitude of the signal at the
            	detector given by equation (23) in~\cite{Owen2010}
			%
			%%%%%%%%%% h_0 equation %%%%%%%%%%
			%
			\begin{equation}
				h_0= \sqrt{\frac{8\pi}{5}}\frac{G}{c^5}\frac{1}{r}\alpha\omega_r^3MR^3\tilde J.
				\label{eq:h0_alpha}
			\end{equation}
            	This expresses the amplitude of the signal in terms of the NS mass
            	$M$,  radius $R$, distance from the observer $r$, a dimensionless
            	parameter $\tilde J$ (1.635$\times$ $10^{-2}$) that depends on the
            	stellar mass distribution and the amplitude of the mode oscillation
            	$\alpha$. We consider the amplitude $\alpha$ in more detail in
            	section~\ref{sec:plausibility}.
            
           	According to \cite{pbr81}, for sufficiently slowly rotating stars, the $r$-mode oscillation
            	frequency $f_r$ is proportional to the spin frequency of the star
           	$f_{\mathrm{spin}}$.  It is expected that the $\ell=m=2$ mode
            	dominates  the GW emission \cite{LockFriedAnd2003}.  Under these assumptions
            	the  $r$-mode frequency
            	is given by,
           	%%%%%%%%%% omega equation %%%%%%%%%%
           	%
            	\begin{equation}\label{eq:rmode_freq}
            		\omega_r \approx \frac{4}{3}\Omega,
            	\end{equation}
            	where $\Omega$ is the angular spin frequency of the star (= 2$\pi$$f_{spin}$) \cite{pbr81}. 
            	There are several sources of uncertainty in this relation.  These include
            	rotational effects~\cite{LindMendOw1999}, which modify the mode
                   frequency by an amount that depends upon the average density, and relativistic 
            	effects~\cite{LockFriedAnd2003}, which introduce fractional corrections of the order 
		of the stellar compactness $M/R$, so that equation (\ref{eq:rmode_freq}) is accurate
		only to the $\sim 10\%$ level.  
            
            	The damping time of the $r$-mode is given by,
			%
			%%%%%%%%%% tau equation %%%%%%%%%%
			%
			\begin{equation}
				\label{eq:tau_sum}
				\frac{1}{\tau_r} = \frac{1}{\tau_{\mathrm{viscosity}}} +
               			 \frac{1}{\tau_{\mathrm{GRR}}},
			\end{equation}
            	where  $\tau_{{\mathrm viscosity}}$ is the dissipation time-scale
            	due to the viscosity in the fluid component of the NS and
            	$\tau_{\mathrm{GRR}}$ is the gravitational radiation reaction
            	time-scale.
            	Following~\cite{Owenetal1998}, the radiation reaction time-scale is
            	a function of the mode frequency,
			%
			%%%%%%%%%% tau_GRR equation %%%%%%%%%%
			%
			\begin{equation}
				\label{eq:GRR}
				\frac{1}{\tau_{\mathrm{GRR}}} =
               			-\frac{32\pi}{15^2}\frac{G}{c^7}\omega_r^6\tilde J M R^4,
			\end{equation}
            	Note that the damping time of the signal $\tau_r$ is a balance between dissipative
            	processes with time-scale $\tau_{\mathrm{viscosity}}>0$ and the
            	radiation reaction time-scale $\tau_{\mathrm{GRR}}<0$; while
            	$\tau_{\mathrm{viscosity}}$ acts to dissipate energy and reduce the
            	oscillation amplitude of the mode, $\tau_{\mathrm{GRR}}$ forces the
            	mode's amplitude to grow due to the emission of gravitational
            	radiation \cite{GravRadia_Lindblom_etal}.
	         In this work we assume that the mode is in the stable regime so that 
            	$\tau_{\mathrm{viscosity}}$$<<$$|\tau_{\mathrm{GRR}}|$, and so the
            	viscosity time-scale dominates in the duration of the signal.
			
            	Levin and Ushomirsky proposed in~\cite{LevinUshutoymodel} the
            	existence of a viscous boundary layer in the crust-core interface of
            	the NS which yields the following estimate of the dissipative
            	time-scale,
		 	%
			%%%%%%%%%% tau_vbl equation %%%%%%%%%%
			%
			\begin{equation}
				\frac{1}{\tau_{\mathrm{viscosity}}}\simeq 0.01
                			s^{-1}\frac{R_{6} ^2
                            		F^{1/2}}{M_{1.4}T_{8}}\frac{\rho_b}{\rho}\bigg(\frac{f_{\rm spin}}
				{\mathrm{kHz}}\bigg)^{1/2}\bigg(\frac{\delta u}{u}\bigg)^2,
				\label{eq:tau_LU}
			\end{equation}
            	where $R_{10}$, $M_{1.4}$, $\rho$ and $\rho_{b}$ are the radius (10
            	Km), mass (1.4 $M_{\odot}$), the density at the crust-core interface, and an estimate
		of this density ($\rho_{\rm b} = 1.5 \times 10^{14}$ g cm$^{-3}$), respectively.
            	$T_8$ is an assumed internal temperature of $10^8$ K in the NS.	The value of the
            	fitting parameter $F$ depends on the dominant scattering processes
            	in the fluid interior \cite{BildstenUsho2000}: 

           		 \begin{equation}
            			F\sim 
            			\begin{cases}
            				1/15 & \text{for electron--proton scattering,} \\
                    			(\rho/\rho_b)^{5/4} & \text{for neutron scattering,} \\
                    			5\rho/\rho_b & \text{for electron scattering.}
            			\end{cases}
            		\end{equation}
 		Here, we assume that neutron-scattering dominates and
            	we have set $\rho$ = $\rho_{b}$.  
            
            	Finally, the ratio $\delta u/u$ measures the fractional velocity mismatch between
            	the crust and the core.  As demonstrated in \cite{LevinUshutoymodel}, for a sufficiently
            	slowly rotating star, the crust does not significantly participate in the r-mode oscillation,
            	so that $\delta u/u \approx 1$.  Just how slow the rotation needs to be for this to be the
            	case depends upon the thickness of the crust, but Figure 1 of \cite{LevinUshutoymodel}
            	indicate that all young glitching pulsars are likely to have $\delta u/u \approx 1$.  Note,
            	however, that in more  rapidly rotating stars $\delta u /u \approx 0.1$, lengthening the
            	viscous decay timescale by a factor of $\sim 100$. 

		%%%%%%%%%% SECTION 3  %%%%%%%%%%
		%
		\section{Detectability}\label{sec:plausibility}
	         The detectability of the GW signal from $r$-mode oscillations
            	excited by a pulsar glitch is determined by the total amount of
            	energy that the glitch deposits in the r-mode, and by the fraction of that
            	energy which is radiated as GWs by the $r$-mode oscillations.

            	Following \cite{Owen2010}, the instantaneous gravitational wave luminosity
            	from a source at distance $r$,  undergoing damped oscillations with
            	initial amplitude $h_0$, angular frequency $\omega$ and damping time
            	$\tau$ is,
			%
			%%%%%%%%%% equation E_GW  %%%%%%%%%%
			%
			\begin{equation}
				\dot E _{\mathrm{GW}} = \frac{c^3}{G}\frac{1}{10}r^2\omega^2(h_0 e^{-t/\tau})^2 .
				\label{eq:Edot}
			\end{equation}  
            	The total time-integrated energy emitted in GWs is, therefore,
			%
			%%%%%%%%%% equation E_GW  %%%%%%%%%%
			%
			\begin{equation}
				\Delta E_{GW} = \frac{c^3}{G}\frac{1}{20}r^2\omega^2h_0^2\tau.
				\label{eq:EGW}
			\end{equation}  
            	The mode energy  is
			%
			%%%%%%%%%% equation mode's energy  %%%%%%%%%%
			%
			\begin{equation}
				\label{eq:En_mode}
				\tilde E = \alpha^2 \Omega^2 M R^2 \tilde J.
			\end{equation} 
            	A rough estimate of the total energy associated with a pulsar glitch
            	is (see e.g. ~\cite{Velapaper2011}),
			%
			%%%%%%%%%% equation glitch energy  %%%%%%%%%%
			%
			\begin{equation}
				\label{eq:Eglitch}
				E_{\mathrm{glitch}} \approx I\Omega^2
                			\left(\frac{\Delta\Omega}{\Omega}\right),
			\end{equation}
            	where $I$ the moment of inertia of the NS, $\Omega$ is the spin
            	frequency and $\Delta \Omega/\Omega$ is the size of the glitch
            	relative to the spin-frequency. An upper limit on the
            	mode amplitude $\alpha$ can be obtained by assuming all of the
            	energy associated with the glitch is channelled into $r$-mode
            	excitation and we find,
			%
			%%%%%%%%%% equation alpha  %%%%%%%%%%
			%
			\begin{equation}
				\label{eq:alpha}
				\alpha = \bigg(\frac{\tilde I}{\tilde J}\bigg)^{1/2}
                			\bigg(\frac{\Delta\Omega}{\Omega}\bigg)^{1/2},
			\end{equation}
            
		{where $\tilde I$ (typically $\approx 0.261$) is a
              	dimensionless parameter dependant on the stellar mass
              	distribution \cite{Owenetal1998}. To evaluate the total power
            	contained in short-duration, narrow-band signals, it is convenient
            	to use the root-sum-squared amplitude:
			%
			%%%%%%%%%% equation hrss definition  %%%%%%%%%%
			%
			\begin{equation}
			            h_{\mathrm{rss}} = \left[\int_0^{\infty} h_+^2(t) +
                            	            h_{\times}^2(t)~{\mathrm
                                        d}t\right]^{\frac{1}{2}} \approx h_0
                                        \tau^{1/2},
				\label{eq:dot_E_GW}
			\end{equation}
            	where the right hand side is the approximate value for the damped
            	sinusoid we consider for our GW signal. Combining equations
		(\ref{eq:h0_alpha}),~(\ref{eq:Eglitch}) and~(\ref{eq:alpha}), the
            	root-sum-squared GW amplitude from $r$-mode excitation in terms of
            	the stellar parameters and the size of the glitch is,
			%
			%%%%%%%%%% equation hrss  %%%%%%%%%%
			%
			\begin{equation}
				\label{eq:hrss_explicit}
				h_{\mathrm{rss}} =
                			\frac{128}{27}\sqrt{\frac{2\pi}{5}}\frac{G}{c^5}\frac{M}{r}(\Omega
                                      R)^3
				\bigg(\tilde I \tilde J\frac{\Delta\Omega}{\Omega} \tau\bigg)^{1/2}.
			\end{equation}

			%
			%%%%%%%%%% FIGURE 1  %%%%%%%%%%
			%  t
			\begin{figure}[h]
				\begin{center}
					\includegraphics[height=80mm]{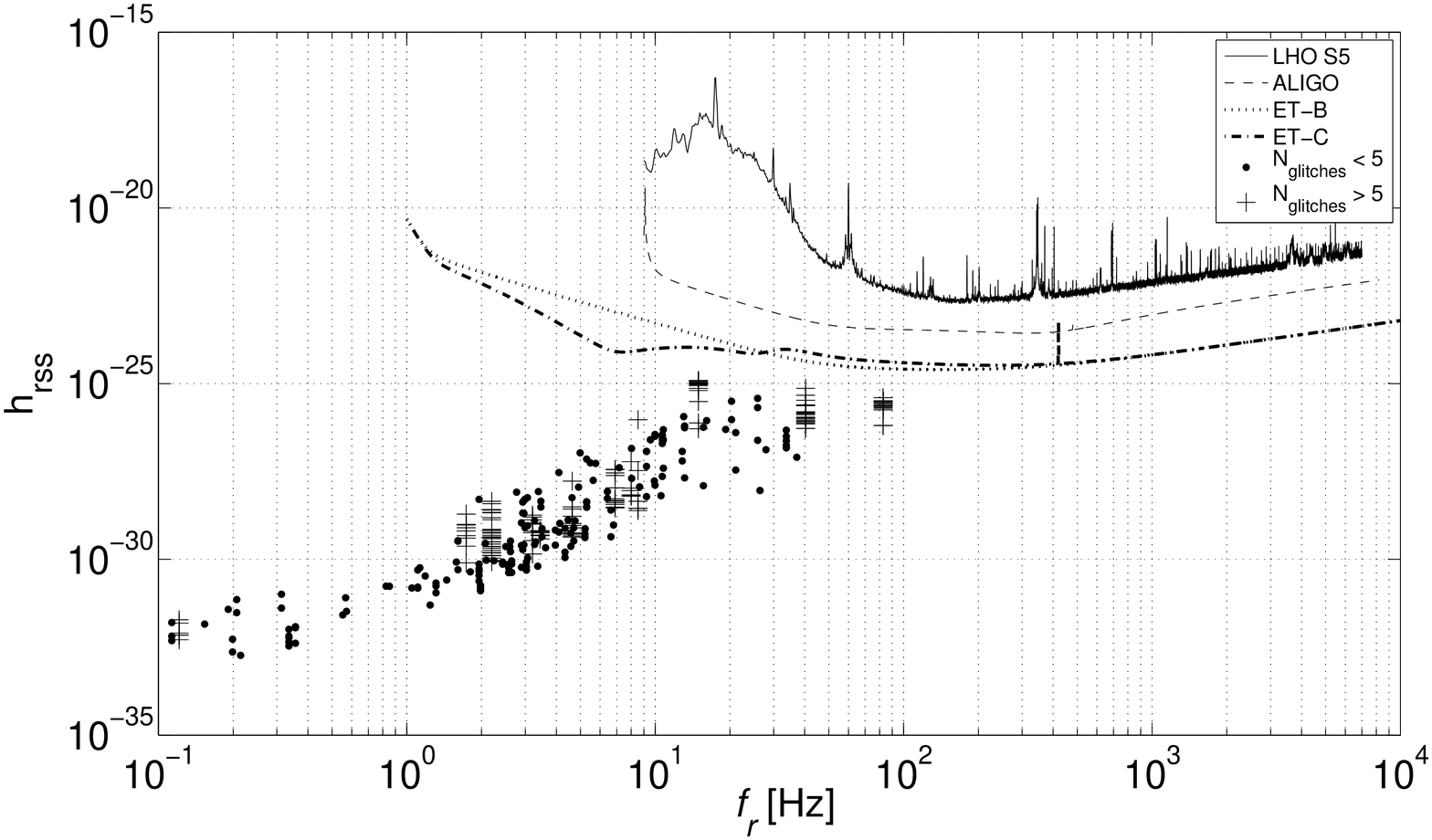}
				\end{center}
                			\caption{\label{fig:detectability}  Sensitivity curves of
             			gravitational waves detectors and upper limit estimates $h_{\mathrm{rss}}$
				 of the gravitational wave signal as a function of the r-mode
             			frequency $f_r = \omega_r/(2\pi)$.}
			\end{figure}
            	In figure~\ref{fig:detectability}, we compare the estimated GW upper limits
            	from glitch-induced $r$-mode excitations in various known
            	pulsars with the noise spectral densities of the initial LIGO
            	instrument during its fifth science run \cite{s5_curve}, the advanced LIGO detector \cite{aligo},\cite{urltocurves}
            	(currently under construction) and the proposed $3^{\mathrm{rd}}$
            	generation Einstein Telescope (ET) \cite{ETcurve}.  We compare the estimates for
            	frequently glitching pulsars (observed to have glitched more than
            	five times, shown in crosses) and all other pulsars, shown in
           	dots.  The glitch observations are compiled from
            	\cite{ATNFdatabase,Melatosetal2007,Yuanetal2010}
            	and~\cite{Espinozaetal2010}. We find that, in
            	this optimistic case, the estimated GW amplitude is a factor $\sim 6$
            	below the noise floor of even ET in its most favorable configuration
            
		 %%%%%%%%%%  SECTION 4  %%%%%%%%%%
				
		\section{Discussion}\label{sec:Discussion}
			
            	We have presented some estimates of the energy associated with
            	$r$-modes excited by pulsar glitches and the corresponding estimated upper limit GW
            	amplitudes.  We see that, even in the most optimistic scenario of energy 
		transfer from the glitch to the mode, the GW amplitude is well below
            	the sensitivity curves of both existing and planned gravitational
		wave detectors.  Specifically, even for the most rapidly
            	spinning pulsar we consider, where the
            	$r$-mode frequency lies near the most
            	sensitive part of the detectors, the expected
            	amplitude is $\sim 4 \times 10^{-26}$.

		However, the strong frequency dependence of the $r$-mode
            	oscillation implies that  rapidly rotating NS have a significantly greater chance of
            	detection in $3^{\mathrm{rd}}$ generation instruments such as ET. In
            	addition to this, it is possible that these faster rotators possess
		a significantly smaller slippage parameter $\delta u/u$ (see
            	section~\ref{sec:signal}), leading to longer duration and more
            	detectable GW signals.  There is clearly some interest in exploring r-mode
		excitation in more rapidly rotating stars. 
            
            	Related to this, we end by noting that, despite the greater frequency of glitches
            	occurring in young pulsars, there is a report of a millisecond
            	pulsar (MSP) glitch in~\cite{Mandaletal2005}.  As well as pushing
            	the expected GW signal closer to the sensitive regions of
            	interferometric GW detectors, higher spin frequencies yield
            	intrinsically stronger signals.  Accreting millisecond pulsars
            	present a particularly tantalising opportunity for future analyses
            	and the GW $r$-mode detectability from such sources is to be
            	considered in future work.
			
		\section*{References}
			\bibliography{ref}
	\end{document}